\documentclass[12pt]{article}
\usepackage[english]{babel}
\usepackage{latexsym}
\usepackage{amsfonts}
\usepackage{epsfig}
\input epsf

\begin{document}
\begin{flushright}
SINP MSU 2004-26/765
\end{flushright}
\begin{center}
{\Large \bf Energy scales in a stabilized brane world}

\vspace{4mm}

Edward E. Boos$^{a,b}$, Yuri S.  Mikhailov$^c$,  Mikhail
N.~Smolyakov$^{a,c}$, Igor P.~Volobuev$^a$\\
\vspace{4mm}
$^a$ Skobeltsyn Institute of Nuclear Physics,
Moscow State University
\\ 119992 Moscow, Russia \\
$^b$ Fermilab, P.O. Box 500, Batavia, IL 60510-0500, USA
\\
$^c$ Physics Department, Moscow State University \\ 119992 Moscow,
Russia \\
\end{center}

\begin{abstract}
Brane world gravity looks different for observers on positive and
negative tension branes. First we consider the  well-known RS1
model with  two branes embedded into the $AdS_5$ space-time and
recall the results on the relations between the energy scales  for
an observer on the negative tension brane, which is supposed to be
"our" brane. Then from the point of view of  this observer we
study  energy scales and masses for the radion and graviton
excitations in a stabilized brane world model. We argue that there
may be several possibilities leading to scales of the order 1-10
$TeV$ or even less for  new physics effects on our brane. In
particular, an  interesting scenario can  arise in the case of a
"symmetric"\ brane world with a nontrivial warp factor in the
bulk, which however takes equal values on both branes.
\end{abstract}

Brane world models with extra space dimensions and different
fields living in the bulk and on the branes present a large number
of intriguing possibilities to solve or to address from a new
viewpoint various problems in particle physics and cosmology.
These are the hierarchy problem,  the proton stability, the
baryogenesis and leptogenesis, the small masses and large mixing
of neutrinos, the  dark matter and dark energy etc. It was
demonstrated that such models can be related to the string
theories and can incorporate sypersymmetry and other possible
symmetries in a natural way. Moreover,  the brane world models
lead to very interesting predictions for experiments at TeV energy
colliders, - Tevatron, LHC, and ILC. \footnote{It is almost
impossible to give references to a very large number of papers in
this field. We refer to the recent this year reviews
\cite{review2004} where original references can be found.}

In brane world scenarios the crucial point is the energy scales
involved in the model, which are usually defined by the
gravitational interaction. This is a reflection  of the fact that
it is gravity that forms the space-time framework for such models.
In the present paper we are going to study the energy scales
rendered by the gravitational interaction in stabilized brane word
scenarios.

We start with considering  the simplest well-known case with  two
branes embedded into the $AdS_5$ space-time, - the RS1 model. It
is based on  the  exact solution for a system of two branes
interacting with gravity in a five-dimensional space-time, which
was found in  paper \cite{Randall:1999ee}. The theory of
linearized gravity interactions based on this solution is called
the Randall-Sundrum model (usually abbreviated as RS1 model), and
it is widely discussed in the literature. An apparent flaw of this
model is the presence of a massless scalar mode called the radion,
whose interactions may contradict the existing experimental data.
This feature of the model reflects the fact that the brane
separation distance is not fixed in it, but remains an arbitrary
parameter.

An approach to curing this flaw was put forward in paper
\cite{Goldberger:1999uk}, where it was shown that a bulk scalar
field with an appropriate interaction can stabilize the brane
separation distance. In this case the radion should acquire a
mass, which is defined by the backreaction of the scalar field on
the metric. Though the latter was not considered in paper
\cite{Goldberger:1999uk}.

An exact solution for a system of two branes interacting with
gravity and a scalar field in a five-dimensional space-time was
found in \cite{DeWolfe:1999cp}. This solution allows one to take
into account the backreaction of the scalar field on the metric
and thus to find the radion mass. Under certain assumptions the
radion mass was computed in paper \cite{Csaki:2000zn}.

Although the solution for the metric in \cite{DeWolfe:1999cp} has
more parameters than the original RS1 solution, it is usually
tacitly implied that the parameters coinciding with those of the
RS1 model take the same values and that the  five- and
four-dimensional energy scales are still related by the formulas
of the RS1 model.

In the present paper we  study this problem in more detail and
show that there are other possibilities of choosing the parameters
in addition to those suggested by the RS1 model. To this end we
reconsider first the relations between the five- and
four-dimensional energy scales in the RS1 model. We follow the
methods and the notations of paper \cite{Randall:1999ee}.

Let us denote the coordinates  of five-dimensional space-time by $
\{ x^M\} \equiv \{x^{\mu},y\}$, $M= 0,1,2,3,4, \, \mu=0,1,2,3$,
the coordinate $x^4 \equiv y, \quad -L\leq y \leq L$
parameterizing the fifth dimension. It forms the orbifold
$S^{1}/Z_{2}$, which is realized as the circle of the
circumference $2L$ with the points $y$ and $-y$ identified (thus,
$y$ is the linear coordinate on the circle, which is related to
the angular coordinate $\phi$ used in \cite{Randall:1999ee} by $y
= \pi \phi$, in particular $L = \pi r_c$). Correspondingly, the
metric $g_{MN}$ satisfies the orbifold symmetry conditions
\begin{eqnarray}
\label{orbifoldsym}
 g_{\mu \nu}(x,- y)=  g_{\mu \nu}(x,  y), \\
 \nonumber
  g_{\mu 4}(x,- y)= - g_{\mu 4}(x,  y), \\ \nonumber
   g_{44}(x,- y)=  g_{44}(x,  y).
\end{eqnarray}
The branes are located at the fixed points of the orbifold, $y=0$
and $y=L$.

The action of the model is
\begin{equation}\label{actionRS}
 S = S_g + S_b,
\end{equation}
where $S_g$ and $S_b$ are given by
\begin{eqnarray}\label{actionsRS}
S_g&=&  \int d^{4}x \int_{-L}^L dy \left(2 M^3
R-\Lambda\right)\sqrt{-g},\\ \nonumber S_b&=& -\lambda_1
\int_{y=0} \sqrt{-\tilde g}  d^{4}x-\lambda_2
\int_{y=L}\sqrt{-\tilde g}   d^{4}x .
\end{eqnarray}
Here $\tilde g_{\mu\nu}$ is the induced metric on the branes and
the subscripts 1 and 2 label the branes. We also note that the
signature of the metric $g_{MN}$ is chosen to be $(-,+,+,+,+)$.
The Randall-Sundrum solution for the  metric is {given by}
\begin{equation}\label{metricrs}
ds^2=  e^{-2\sigma(y)}\eta_{\mu\nu}  {dx^\mu dx^\nu} +
  dy^2 \equiv \gamma_{MN}(y) {dx^Mdx^N} ,
\end{equation}
where $\eta_{\mu\nu}$ is the Minkowski metric and {the function}
\begin{equation}\label{fsigma}
\sigma(y) = k|y|+ c
\end{equation}
in the interval $-L \leq y \leq L$. It is well known that the
constant $c$  in the definition of $\sigma$ can be eliminated by
rescaling the coordinates $\{x^\mu\}$ in (\ref{metricrs}).
Nevertheless, we will keep it, because it will be useful for our
considerations.

The parameter  $k$ is positive and has the dimension of mass, the
bulk cosmological constant $\Lambda$ and the brane tensions $
\lambda_{1,2}$ are  defined  by the fine-tuning conditions:
\begin{equation}\label{cc}
\Lambda = -24M^3 k^2, \quad \lambda_1 =-\lambda_2= 24M^3 k.
\end{equation}
We see that the brane at $y=0$ has a positive energy density,
whereas the  brane at $y=L$ has a negative one.

The linearized theory in the RS1 background is obtained by
representing the metric as
\begin{equation}\label{metricpar}
g_{MN}= \gamma_{MN}+  h_{MN},
\end{equation}
substituting it into action (\ref{actionRS}) and keeping only the
terms quadratic  in $ h_{MN}$ \cite{Boos:2002ik}. The
corresponding equations of motion can be decoupled
\cite{Boos:2002ik} to give an equation for a transverse-traceless
tensor field $b_{\mu\nu}(x,y)$, which describes the graviton and
its massive excitations, and a massless radion scalar field
$\phi(x)$. The equation for the field $b_{\mu\nu}(x,y)$ looks like
\begin{equation}\label{eqb}
\frac{1}{2}(e^{2\sigma(y)}\Box{b_{\mu\nu}}+\frac
{\partial^2{b_{\mu\nu}}}{\partial y
^2})-b_{\mu\nu}(2(\sigma')^2-\sigma'')=0
\end{equation}
and defines the mass spectrum of the tensor particles (here and
below $\Box =\eta^{\mu\nu}\partial_\mu\partial_\nu$ is the
D'Alembert operator and the prime denotes the derivative with
respect to $y$). It is a matter of simple calculation to check
that the massless modes of this equation have the form
\begin{equation}\label{zerom}
b_{\mu\nu}(x,y) =  e^{-2\sigma(y)}\bar h_{\mu\nu}(x).\
\end{equation}
In order to derive a relation between the four- and five
dimensional energy scales, following paper \cite{Randall:1999ee}
we  consider the metric  only with the zero mode excitations
\begin{eqnarray}\label{metric0}ds^2=
e^{-2\sigma(y)}(\eta_{\mu\nu}+\bar h_{\mu\nu}(x))  {dx^\mu dx^\nu}
+dy^2=e^{-2\sigma(y)}\bar g_{\mu\nu}(x) {dx^\mu dx^\nu} +
dy^2,\end{eqnarray} where we introduced the notation
\begin{equation}\label{barg}
\bar g_{\mu\nu}(x) =\eta_{\mu\nu}+\bar h_{\mu\nu}(x).
\end{equation}

If we substitute this metric into action (\ref{actionRS}), we  get
the following result:
\begin{equation}\label{action_g}
S=  2 M^3\int_{-L}^L e^{-2\sigma(y)}dy \int d^{4}x R_4(\bar
g)\sqrt{- \bar g},
\end{equation}
$R_4(\bar g)$ being the four-dimensional scalar curvature.
Integrating in (\ref{action_g}) over the extra dimension, we get
an effective action
\begin{equation}\label{action_eff}
S_{eff}=  2 M^3 e^{-2c}\, \frac{1-e^{-2kL}}{k}\int d^{4}x R_4(\bar
g)\sqrt{- \bar g}.
\end{equation}
 Thus, we see that the reduced action is the standard gravitational action, and
therefore the field $\bar h_{\mu\nu}(x)$ is the graviton field.

The expression in front of the integral is  the coupling constants
of the graviton field to matter on the branes. Since we have two
branes, we have to find out explicitly, to which brane this
coupling constant corresponds.

The coupling of gravity to matter on any brane is given by
\begin{equation}\label{coupling}
\int d^{4}x \sqrt{- \tilde g} L(\phi, \tilde g),
\end{equation}
where $\tilde g$ is the metric induced on the brane and $\phi$
denotes an arbitrary set of fields. An important point is that in
order to get the canonical energy-momentum tensor,  the Lagrangian
$L$ must have  the canonically normalized kinetic terms. For
example, in the case of $\phi$ being a scalar field  it has  the
form $-\frac{1}{2}\tilde
g^{\mu\nu}\partial_\mu\phi\partial_\nu\phi$.

Linearizing  interaction (\ref{coupling}), we get
\begin{eqnarray}\label{coupling_l}\int d^{4}x \sqrt{- \gamma}
\left(\frac{\delta L}{\delta \gamma_{\mu\nu}} - \frac{1}{2}
\gamma^{\mu\nu} L\right) \delta \tilde g_{\mu\nu} =\frac{1}{2}\int
d^{4}x \sqrt{- \gamma}e^{-2\sigma}\bar h_{\mu\nu} T^{\mu\nu} =\\
\nonumber = \frac{1}{2}\int d^{4}x e^{-2\sigma} \bar h_{\mu\nu}
T_{\rho\sigma}\eta^{\mu\rho}\eta^{\nu\sigma}.
\end{eqnarray}

Let us first consider the interaction with matter on the positive
tension brane. Putting $y=0$ in (\ref{fsigma}) and
(\ref{coupling_l}), we get
\begin{equation}\label{metric_0}
ds^2=  e^{-2c}\eta_{\mu\nu}  {dx^\mu dx^\nu} +   dy^2 =
\gamma_{\mu\nu}(0) {dx^\mu dx^\nu} +   dy^2,
\end{equation}
and the interaction with matter of the form
\begin{equation}\label{coupling_0}
\frac{1}{2}\int d^{4}x e^{-2c}\bar h_{\mu\nu}
T_{\rho\sigma}\eta^{\mu\rho}\eta^{\nu\sigma}.
\end{equation}
We recall (see \cite{LL}) that the coordinates are called
Galilean, if $\gamma_{\mu\nu}= diag(-1,1,1,1)$. Though physics
does not depend on the choice of the  coordinates,  this is the
preferred coordinate system, in which we interpret  experimental
results and where the space coordinates are measured by a "ruler",
the time is measured by a "clock" and all the  physical
observables  are measured in the standard  units (cm, GeV, etc.).
Equation (\ref{metric_0}) implies that the coordinates $\{x^\mu\}$
in (\ref{metricrs}) are Galilean on the  brane at $0$ only if
$c=0$. In this case equation(\ref{action_eff}) gives the
well-known relation between the energy scale $M_P$ on the positive
tension brane and in the bulk \cite{Randall:1999ee}:
\begin{equation}\label{rel1}
M^2_P=  M^3\frac{1-e^{-2kL}}{k}.
\end{equation}
Since for an observer on the positive tension brane $M_P$ is equal
to the four-dimensional gravitational energy scale on this brane,
he has to conclude that $M\sim k$ are also of the same order.
Thus, for an observer on the positive tension brane the
fundamental five-dimensional energy scale is of the order of the
four-dimensional gravitational energy scale on this brane.

Now let us  consider the interaction with matter on the negative
tension brane. Putting $y=L$ in (\ref{fsigma}) and
(\ref{coupling_l}), we get
\begin{equation}\label{metric_L}ds^2=  e^{-2(kL+c)}\eta_{\mu\nu}  {dx^\mu
dx^\nu} +   dy^2= \gamma_{\mu\nu}(L) {dx^\mu dx^\nu} + dy^2,
\end{equation}
and the interaction Lagrangian
\begin{equation}\label{coupling_L}\frac{1}{2}\int d^{4}x
e^{-2(kL+c)} \bar h_{\mu\nu}
T_{\rho\sigma}\eta^{\mu\rho}\eta^{\nu\sigma}.
\end{equation}
Now equation (\ref{metric_L}) gives that the coordinates
$\{x^\mu\}$ in (\ref{metricrs}) are Galilean on the  brane at $L$
only if $c=-kL$. Substituting this value into  equation
(\ref{action_eff}) gives the  relation between the energy scale
$M_N$ on the negative tension brane and in the bulk
\cite{Boos:2002ik,Rubakov:2001kp,Giudice:2004mg}:
\begin{equation}\label{rel2}
M^2_N=  M^3\frac{e^{2kL}-1}{k}.
\end{equation}
For an observer on the negative tension brane $M_N=M_{PL}$, which
can be achieved by taking $M\sim k\sim 1 TeV$ and $kL \simeq 35$.
Thus, for an observer on the negative tension brane the physical
picture is quite different from that of an observer on the
positive tension brane. The fundamental five-dimensional energy
scale can lie in the TeV region, whereas the four-dimensional
energy scale, explaining the weakness of the gravitational
interaction, arises due to the warp factor in the metric.

Thus, we see that the gravity in the bulk looks different, when
viewed from different branes. In the case of the negative tension
brane one can speak of a "bottom-up" picture, when a large energy
scale arises from a small one due to the warp factor. In other
words, the fundamental energy scale can be of the TeV order and
the four-dimensional Planck scale on our brane appears as an
effective scale $M_{PL} = M e^{kL}$, which explains the observed
weakness of Newtonian gravity.

Now let us consider a  stabilized brane world  model. To be
specific, we take the model admitting an exact solution  found  in
\cite{DeWolfe:1999cp}. Its action can be written as
\begin{equation}\label{actionDW}
S = S_g + S_\phi,
\end{equation}
where $S_g$ and $S_\phi$ are given by
\begin{eqnarray}\label{actionsDW}
S_g&=& 2 M^3\int d^{4}x \int_{-L}^L dy  R\sqrt{-g},\\ \nonumber
S_\phi &=& -\int d^{4}x \int_{-L}^L dy \left(\frac{1}{2}
g^{\mu\nu}\partial_\mu\phi\partial_\nu\phi+V(\phi)\right)\sqrt{-g}
-\\ \nonumber & -&\int_{y=0} \sqrt{-\tilde g}\lambda_1(\phi)
d^{4}x -\int_{y=L}\sqrt{-\tilde g}\lambda_2(\phi)    d^{4}x .
\end{eqnarray}
Here $V(\phi)$ is a bulk scalar field potential and
$\lambda_{1,2}(\phi)$ are brane   scalar field potentials. They
will be specified later.

The standard ansatz  for the  metric and the scalar field, which
preserves the Poincar\'e invariance in any four-dimensional
subspace $y=const$ looks like
\begin{eqnarray}\label{metricDW}
ds^2&=&  e^{-2A(y)}\eta_{\mu\nu}  {dx^\mu  dx^\nu} +  dy^2 \equiv
\gamma_{MN}(y)dx^M dx^N, \\ \phi &=& \phi(y).
\end{eqnarray}
If one substitutes this ansatz into action  (\ref{actionDW}), one
gets a rather complicated system of nonlinear differential
equations for functions $A(y), \phi(y)$ \cite{DeWolfe:1999cp}. If
the bulk potential $V(\phi)$ can be represented as $$
V(\phi)=\frac{1}{8} \left(\frac{\delta W}{\delta\phi}\right)^2-
\frac{{1}}{24 M^3}W^2(\phi),
 $$
the original system in the bulk is equivalent to a pair of
equations
\begin{equation}
\phi^\prime(y) =\frac{1}{2} \frac{\delta W}{\delta\phi}, \quad
A^\prime (y) = \frac{{1}}{24 M^3}W(\phi),
\end{equation}
which can be solved exactly for an appropriate choice of the
function $W(\phi)$. To get a solution of the complete system, one
has to fine-tune the brane scalar field potentials, which is
similar to fine-tuning the cosmological constant and the brane
tensions in the RS1 model.

To obtain a quartic potential $V(\phi)$, the function $W$ can be
chosen as
\begin{equation}\label{fW}
W(\phi) = {24}{ M^3}k - u \phi^2,
\end{equation}
where the parameters $k$ and $u$ have the dimension of mass. Then
the fine-tuned brane potentials must be of the form
\begin{eqnarray}\label{branepot}
\lambda_1(\phi) &=& W(\phi_0)+ W^\prime (\phi_0)(\phi-\phi_0) +
\beta_1^2(\phi-\phi_0)^2, \\ \nonumber \lambda_2(\phi) &=&
-W(\phi_L)- W^\prime (\phi_L)(\phi-\phi_L) +
\beta_2^2(\phi-\phi_L)^2.
\end{eqnarray}
The parameter $k$ is, in fact, the standard parameter of the RS1
model, which is related to the bulk cosmological constant $\Lambda
= V(0)$ by the same formula, as in (\ref{cc}). Thus, the
parameters of the model  are  $k, u, \phi_0, \phi_L, \beta_{1,2}$,
as well as the fundamental five-dimensional gravitational scale
$M$. When the former parameters are made dimensionless with the
help of  $M$, they should be of the order $O(1)$, so that there is
no hierarchical difference in the  parameters.

For such a choice of the potentials, the solutions for  functions
$A(y), \phi(y)$ look like
\begin{eqnarray}\label{sols}
\phi(y) &=& \phi_0 e^{-u|y|}, \\ \nonumber A(y) &=& k|y| +
\frac{\phi_0^2}{48 M^3}e^{-2u|y|} +c.
\end{eqnarray}
In the sequel we will denote
\begin{equation}\label{constb}
b=\frac{\phi_0^2}{48 M^3},
\end{equation}
and write $A(y)$ as
\begin{equation}\label{fA}
A(y) = k|y| + b e^{-2u|y|} +c.
\end{equation}
Here again we keep the constant $c$, which can be removed by a
rescaling of the coordinates $\{x^\mu\}$ in metric
(\ref{metricDW}). Now in order to have Galilean coordinates one
has to take $c = -b$ for the positive tension brane and $c = -kL -
b e^{-2uL}$ for the negative tension brane.

The separation distance in defined by the equation
\begin{equation}\label{sepL}
L= \frac{1}{u}\ln \frac{\phi_0}{\phi_L}
\end{equation}
and,  therefore,  it is stabilized. Keeping this in mind, it is
convenient to use $L$ instead of $\phi_L$ as a parameter of the
theory.

Now the linearized theory is obtained by representing the metric
and the scalar field as
\begin{eqnarray}\label{metricparDW}
g_{MN}(x,y)&=& \gamma_{MN}(y) +  h_{MN}(x,y), \\ \phi(x,y) &=&
\phi_0 e^{-u|y|} +  f(x,y),
\end{eqnarray}
substituting this representation into action (\ref{actionDW}) and
keeping the terms of the second order  in $h_{MN}$ and $f$. The
corresponding equations of motion are:
\begin{enumerate}
\item $\mu\nu$-component
\begin{eqnarray}\nonumber
&\frac{1}{2}(\partial_\sigma{\partial^\sigma{h_{\mu\nu}}}-\partial_\mu
{\partial^\sigma{h_{\sigma\nu}}}-\partial_\nu{\partial^\sigma{h_{\sigma\mu}}}
+\partial_4{\partial_4{h_{\mu\nu}}})+\frac{1}{2}\partial_\mu{\partial_
\nu{\tilde{h}}}+\frac{1}{2}\partial_\mu{\partial_\nu{h_{44}}}&-\\
\nonumber &-\frac{1}{2}\partial_4{
(\partial_\mu{h_{4\nu}}+\partial_\nu{h_{4\mu}})}
+A'(\partial_\mu{h_{4\nu}}+
\partial_\nu{h_{4\mu}})
+\frac{1}{2}\gamma_{\mu\nu}\biggl(-\partial_4 {\partial_4
{\tilde{h}}} -  &\\ \nonumber
&-\partial_{\sigma}{\partial^{\sigma}
{h_{44}}}-\partial_\sigma{\partial^\sigma
{\tilde{h}}}+4A'\partial_4{\tilde{h}}-3A'\partial_4{h_{44}}
+\partial_{\sigma}{\partial_{\tau}{h^{\sigma\tau}}}+
2\partial^{\sigma}{\partial_4{h_{\sigma 4}}}- &\\ \nonumber
&-4A'\partial^\sigma{h_{4 \sigma}}
\biggr)-h_{\mu\nu}(2A'^2-A'')+\frac{3}{2}h_{44}\gamma_{\mu\nu}(4A'^2-A'')-&\\\label{munu}
&-
\frac{1}{4M^3}\left(\gamma_{\mu\nu}f'\phi'-f\gamma_{\mu\nu}(4A'\phi'-\phi'')
\right)=0; &
\end{eqnarray}
\item $\mu 4$-component
\begin{eqnarray}\label{hm4}
\partial_4(\partial_\mu\tilde{h}-\partial^{\nu}
h_{\mu\nu})+\partial^\nu(\partial_\nu h_{\mu 4 }-\partial_\mu
h_{\nu 4})+3A'\partial_{\mu} h_{44}+\\ \nonumber +
\frac{1}{2M^3}\partial_\mu f\phi'=0;
\end{eqnarray}
\item $44$-component
\begin{eqnarray}\label{44eq}
\partial^\mu(\partial^\nu h_{\mu\nu}-\partial_{\mu}
\tilde{h})-6A'\partial^{\mu} h_{\mu4} +3A'\partial_4{\tilde{h}}+
\\ \nonumber + \frac{1}{2M^3}\left(- h_{44}V-f\frac{\delta
V}{\delta\phi}+f'\phi'\right)=0;
\end{eqnarray}
\item equation for the field $f$
\begin{eqnarray}\nonumber
&h_{44}\left(
\frac{\delta\lambda_1}{\delta\phi}\delta(y)+\frac{\delta\lambda_2}
{\delta\phi}\delta(y-L)\right)-2h_{44}(-4A'\phi'+\phi'')+&\\
\nonumber
&+\phi'\partial_4\tilde{h}-\phi'\partial_4h_{44}-8A'\partial_4f
+2\partial_M{\partial^M{f}}-2\phi'\partial^{\mu}h_{\mu 4}-&\\
\label{feq} &-2f\left(\frac{\delta^2 V }{\delta
\phi^2}+\frac{\delta^2 \lambda_1 }{\delta
\phi^2}\delta(y)+\frac{\delta^2 \lambda_2 }{\delta
\phi^2}\delta(y-L) \right)=0.&
\end{eqnarray}
\end{enumerate}
Here and below $\phi$ stands for the background solution
(\ref{sols}) and $\tilde h = \gamma^{\mu\nu}h_{\mu\nu}.$ We will
also use the following auxiliary equation, which is obtained by
contracting the indices in the $\mu\nu$-equation:
\begin{eqnarray}\label{cont}\nonumber
&\partial^\mu \partial^\nu h_{\mu\nu}- \partial^\mu \partial_\mu
(\tilde{h}+\frac{3}{2}h_{44})-
6A'\partial_4(h_{44}-\tilde{h})-\frac{3}{2}\partial_4
\partial_4\tilde{h}-6A'\partial^\mu h_{\mu 4}+&\\
&+3\partial^{\mu}
\partial_4 h_{\mu 4}+6h_{44}(4A'^2-A'')
-\frac{1}{M^3}(f'\phi'+f(-4A'\phi'+\phi''))=0.&
\end{eqnarray}

These equations are invariant under the gauge transformations
\begin{eqnarray}\label{kal}\nonumber
&h_{\mu\nu}'(x,y)=h_{\mu\nu}(x,y)-(\partial_\mu{\xi_\nu}+\partial_\nu{\xi_\mu}
-2\gamma_{\mu\nu}\partial_4{A}\xi_4),&\\ \nonumber
&h_{\mu4}'(x,y)=h_{\mu4}(x,y)-(\partial_\mu{\xi_4}+\partial_4{\xi_\mu}
+2\partial_4{A}\xi_\mu),&\\
&h_{44}'(x,y)=h_{44}(x,y)-2\partial_4{\xi_4},&\\
&f'(x,y)=f(x,y)-\partial_4\phi\xi_4, \quad
\partial_4\equiv{\partial}/{\partial{y}} ,
\end{eqnarray}
provided ${\xi_M(x,y)}$ satisfy the orbifold symmetry conditions
$$ \xi_{\mu}(x,-y)= \xi_{\mu}(x,y), \quad \xi_4(x,-y)= -
\xi_4(x,y). $$ These gauge transformations are a generalization of
the gauge transformations in the unstabilized RS1 model
\cite{Boos:2002ik,Rubakov:2001kp}. We will use them to isolate the
physical degrees of freedom of the fields $h_{MN}$ and $f$.

The gauge transformations with function $\xi_4$ allow one to
impose the gauge condition
\begin{equation}
\label{uf}(e^{-2A}h_{44})'-\frac{1}{3 M^3}e^{-2A}\phi'f=0.
\end{equation}
This relation was obtained in \cite{Csaki:2000zn} from the
equation for $\mu 4$-component, in which only the scalar degrees
of freedom were retained.

Similar to the case of the unstabilized RS1 model,  the gauge
transformations with functions $\xi_\mu$ allow one to impose the
gauge $h_{\mu4}(x,y)=0$. After which there remain the gauge
transformations satisfying
\begin{equation}\label{restr}
\partial_4({e^{2A}\xi_\mu})=0.
\end{equation}

Thus, we  can use the gauge
\begin{eqnarray}\label{gauge}
(e^{-2A}h_{44})'-\frac{1}{3 M^3}e^{-2A}\phi'f=0, \\ \nonumber
h_{\mu 4} =0.
\end{eqnarray}

Next we  represent the gravitational field as
\begin{equation}\label{hb}
h_{\mu\nu}=b_{\mu\nu}+\frac{1}{4}\gamma_{\mu\nu}\tilde{h},
\end{equation}
with  $b_{\mu\nu}$ being a traceless tensor field
($\gamma^{\mu\nu}b_{\mu\nu}=0$).

Substituting gauge conditions (\ref{gauge}) and representation
(\ref{hb}) into the $\mu4$-equation and into  contracted
$\mu\nu$-equation  (\ref{cont}), we get:
\begin{eqnarray} \label{constraintg}
&-\partial_4 (\partial^\nu b_{\mu\nu} ) + \frac{3}{4}\partial_\mu
\partial_4(\tilde h + 2 h_{44}) =0,& \\ \nonumber
&\partial^\mu \partial^\nu b_{\mu\nu} -\frac{3}{4} \partial_\rho
\partial^\rho \tilde h -\frac{3}{2} \partial_\rho \partial^\rho
h_{44} -\frac{3}{2} \frac{\partial^2} { \partial y^2} \tilde h
+&\\ \label{contg} &+ 6A' \partial_4\tilde h -3\frac{\partial^2}{
\partial y^2} h_{44} + 12 A'\partial_4 h_{44}=0. &
\end{eqnarray}

Equation (\ref{constraintg}) suggest the substitution $\tilde h =
- 2 h_{44}$, which allows one to decouple the equations for the
fields $b_{\mu\nu}, h_{44}$ and  $f$. Really, as a result of this
substitution equations (\ref{constraintg}), (\ref{contg}) take the
form
\begin{eqnarray}\label{db}
\partial_4 (\partial^\nu b_{\mu\nu} )  =0, \\
\partial^\mu \partial^\nu b_{\mu\nu} =0.
\end{eqnarray}

It is not difficult to check that the residual gauge
transformations (\ref{restr}) are sufficient to impose the gauge
$$
\partial^\nu{ b_{\mu\nu}}=0,
$$ in which the former equations are satisfied identically.

Thus, in what follows we will be working in the gauge
\begin{eqnarray}\label{compl_gauge}
(e^{-2A}h_{44})'-\frac{1}{3 M^3} e^{-2A}\phi'f=0, \\ \nonumber
h_{\mu 4} =0,\\ \nonumber\partial^\nu{ b_{\mu\nu}}=0,
\end{eqnarray}
 the residual gauge transformations now being
\begin{equation}\label{ok1}
\xi_\mu=e^{-2A}\epsilon_\mu(x),\qquad
\partial^\nu\epsilon_\nu(x)=0,\qquad \partial^\mu{\partial_\mu{\epsilon_\nu}}=0.
\end{equation}

Obviously, after the substitution  $\tilde h = - 2 h_{44}$
contracted $\mu\nu$-equation  (\ref{cont}) and the $\mu4$-equation
are satisfied identically in this gauge. Equation (\ref{munu}) for
the  $\mu\nu$-component reduces to  an equation for a
transverse-traceless tensor field $b_{\mu\nu}(x,y)$:
\begin{equation}\label{ub}
\frac{1}{2}(e^{2A(y)}\Box
{b_{\mu\nu}}+\frac{\partial^2{b_{\mu\nu}}}{\partial
y^2})-b_{\mu\nu}\left(2(A')^2- A''\right)=0.
\end{equation}
This equation  is absolutely analogous to equation (\ref{eqb}) in
the unstabilized RS1 model.

Scalar field equations follow from equations (\ref{44eq}) and
(\ref{feq}). It turns out that these equations are equivalent in
the bulk and look much simpler, when written in terms of a new
function $g$, which is related to $h_{44}$ by $h_{44}(x,y) =
e^{2A(y)}g(x,y)$:
\begin{equation}\label{ug}
e^{2A(y)}\Box{g}+2A'g' -2\frac{\phi''}{\phi'}g'- \frac{ u^2
\phi^2}{6 M^3}  g + g''=0.
\end{equation}
This  equation  is similar to the one studied in
\cite{Csaki:2000zn}. The boundary conditions for this equation
follow from (\ref{44eq}) and (\ref{feq}) and look like
\begin{eqnarray}\label{bc1}
(\beta_1^2 +u)g' + \partial_\mu{\partial^\mu{g}}|_{y=+0} =0 ,\\
\label{bc2} (\beta_2^2 -u)g' -
\partial_\mu{\partial^\mu{g}}|_{y=L-0} =0.
\end{eqnarray}

Again it is easy to check that the massless mode of equation
(\ref{ub}) is given by
\begin{equation}\label{zeromDW}
b_{\mu\nu}(x,y) = e^{-2A(y)}\bar h_{\mu\nu}(x).
\end{equation}
Residual gauge transformations (\ref{ok1}) imply that the massless
mode has only two physical  degrees of freedom. Thus, the metric
including only the massless tensor mode looks like
\begin{eqnarray}\label{metric0DW}
ds^2=  e^{-2A(y)}(\eta_{\mu\nu}+\bar  h_{\mu\nu}(x))  {dx^\mu
dx^\nu} + dy^2\equiv \\ \nonumber \equiv e^{-2A(y)}\bar
g_{\mu\nu}(x) {dx^\mu dx^\nu} +   dy^2,
\end{eqnarray}
$\bar g_{\mu\nu}(x)$ being given by (\ref{barg}).

Substituting this metric and the background field $\phi(y) =
\phi_0 e^{-u|y|}$ into action (\ref{actionDW}), we arrive at the
following expression:
\begin{equation}\label{action_gDW}S=  2 M^3\int_{-L}^L e^{-2A(y)}dy
\int d^{4}x R_4(\bar g)\sqrt{- \bar g},
\end{equation}
the notations being the same as in (\ref{action_g}). Integrating
in the latter action over $y$, we get an effective
four-dimensional action
\begin{equation}\label{action_effDW}
S_{eff}=  2 M_4^2\int d^{4}x R_4(\bar g)\sqrt{- \bar g}.
\end{equation}
The relation between the energy scales is now defined by the
integral
\begin{equation}\label{intDW}
M_4^2={M^3} e^{-2c} \int_{-L}^L dy e^{-2k|y| -2b e^{-2u|y|}} \, ,
\end{equation}
which is explicitly given  by
\begin{equation}\label{couplingDW}
M_4^2=\frac{M^3}{u}e^{-2c}(2b)^{-\frac{k}{u}}
\left\{\gamma\left(\frac{k}{u},2b\right) -
\gamma\left(\frac{k}{u},2be^{-2uL}\right)\right\}.
\end{equation}
Here  $b$ is  defined in (\ref{constb}) and $\gamma$ is the
incomplete gamma function.

Again we got the standard gravitational action and therefore the
field  $\bar h_{\mu\nu}(x)$ is the graviton field. Its coupling to
matter on  the positive tension brane is defined by expression
(\ref{couplingDW}) with  $c=-b$ and the coupling to matter on  the
negative tension brane is defined by the same expression with  $c
= -kL -be^{-2uL}$. Using the standard formulas for the incomplete
gamma function, one can easily check that in the limit
$b\rightarrow 0$  or $u\rightarrow  \infty$ the former expression
goes into (\ref{rel1}), and the latter into (\ref{rel2})
respectively.

One can analyse  relation (\ref{couplingDW}) in the case of
physically motivated values of the parameters involved, but it
turns out to be more convenient to use the integral representation
(\ref{intDW}) rather than the explicit formula in terms of
incomplete gamma functions.
 Similar questions have been addressed in \cite{Csaki:2000zn,Barger:2000wj};
 moreover,  in the latter paper the cosmological aspects of the stabilized warped brane
 models have been discussed.

1. The first obvious approximation is $b=\frac{\phi_0^2}{48
M^3}\ll1 $, i.e. the backreaction of the scalar field on the RS
metric is assumed to be small. This approximation was studied in
\cite{Csaki:2000zn}. Essentially it means that one can drop the
term with $b$ in (\ref{fA}) and retain it only in  equation
(\ref{ug}) for the scalar field. The boundary conditions for this
field being rather complicated, in \cite{Csaki:2000zn} it was
suggested to use the limit $\beta_{1,2} \rightarrow \infty$. In
this limit (\ref{bc1}), (\ref{bc2}) reduce to
\begin{equation}\label{bc3}
g'|_{y=0}= g'|_{y=L} =0.
\end{equation}
The relations between the five- and four-dimensional energy scales
are given by the same formulas (\ref{rel1}) and (\ref{rel2}) with
the same values of the parameters $M$ and $k$. The radion mass,
i.e. the mass of the lowest scalar excitation, for an observer on
the negative tension brane can be  estimated as
\cite{Csaki:2000zn}
\begin{equation}\label{mrad}
m^2_{rad}= 32bu^2 e^{-2uL}
\end{equation}
Recall that on the negative tension brane  $M$, $k$ and  $u$ can
be of the order of 1 TeV. We also must have $kL \sim 35$ in order
to get the correct value of Newton's constant. This means that to
have the radion mass in the range of 100 GeV the ratio $u/k$  has
to be  small. This observation suggests another approximation,
namely $uL\ll1$.

2. Assuming  $uL\ll1$, but keeping $b$ arbitrary, we can expand
the exponential in (\ref{fA}) to the first order in $uL$ and get
\begin{equation}\label{fAap}
A(y) \simeq (k-2bu)|y| +c = \tilde k |y| +c, \quad \tilde k
=(k-2bu).
\end{equation}
Here we restrict ourselves to considering  only the negative
tension brane, where  our world is supposed to be. In this case
the relation between the fundamental five-dimensional mass $M$ and
the Planck mass looks like
\begin{equation}\label{rel2a}
M^2_{Pl}=  M^3\frac{e^{2\tilde kL }-1}{\tilde k}.
\end{equation}
To estimate the  radion mass we have to solve  equation (\ref{ug})
in this approximation. In so doing we also restrict ourselves to
boundary conditions (\ref{bc3}), which were used in
\cite{Csaki:2000zn}.

Passing to a new variable
\begin{equation}
z = \frac{m}{\tilde k} e^{\tilde k y - \tilde k L}
\end{equation}
brings  equation (\ref{ug}) to the form of the Bessel equation, an
approximate solution with boundary condition (\ref{bc3}) at $y=0$
being, up to normalization,
\begin{eqnarray}\label{fradap}
g(z) &=& z^{-(1+\frac{u}{\tilde k})} J_{\alpha} (z),
 \\ \label{alpha}
\alpha &=& \sqrt{(1 + \frac{u}{\tilde k})^2 + 16b\frac{u^2}{\tilde
k^2}}.
\end{eqnarray}

The spectrum of scalar excitations is defined by boundary
condition (\ref{bc3}) at $y=L$, which gives for the lowest radion
mass
\begin{equation}\label{mradap}
m^2_{rad}= \frac{4 \tilde k^2 (\alpha + 1)(\alpha - 1 -
\frac{u}{\tilde k}) }{(\alpha + 1 - \frac{u}{\tilde k})} .
\end{equation}
This equation defines the radion mass for arbitrary values of $b$.
If $b\frac{u^2}{\tilde k^2}\ll1$, we can expand the square root in
(\ref{alpha}) to the leading order in this variable, which gives
\begin{equation}\label{mradap1}
m^2_{rad}= 32bu^2 .
\end{equation}
This obviously coincides with (\ref{mrad}) in the approximation
under consideration.

\begin{figure}
\epsfig{figure=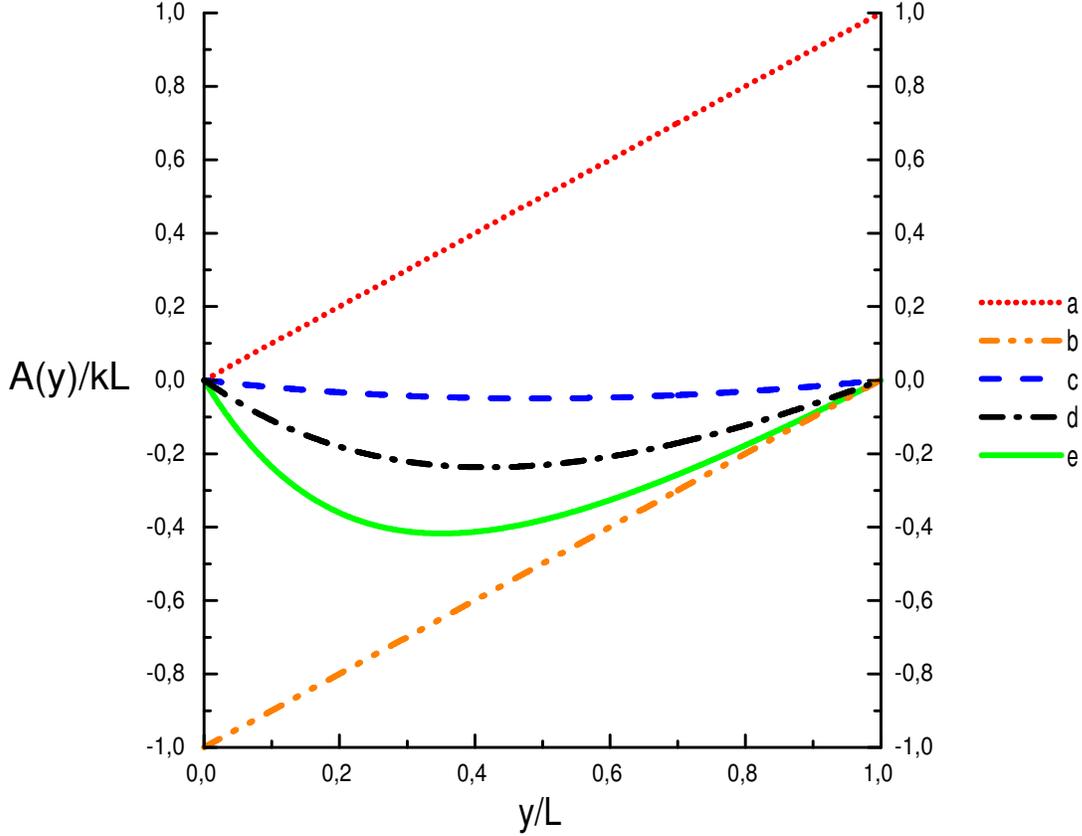,width=16cm} \caption{Plots of
$\frac{A(y)}{kL}$ for different cases: a - RS1 model (observer on
the positive tension brane); b - RS1 model (observer on the
negative tension brane); c - "symmetric" case with $w=0.2$; d -
"symmetric" case with $w=1$; e - "symmetric" case with $w=2$.}
\label{fig}
\end{figure}

3. The discussed scenarios are nevertheless
 very similar to the original RS1 model,
where the four-dimensional  worlds on different branes have
hierarchically different energy scales. For certain values of
parameters the stabilized model can lead to the same energy scale
on both branes. This situation is described by equation
\begin{equation}\label{sym}
A(0) =A(L).
\end{equation}
From equation (\ref{sym}) one can express the b parameter in the
following form
\begin{equation}\label{b_sym}
b = \frac{kL}{1 - e^{-2uL}}.
\end{equation}
Now the function $A(y)$ (\ref{fA}) looks as follows
\begin{equation}\label{A_sym}
A(y) = kL f(\xi),\ f(\xi) = \left\{ \xi - \frac{1 - e^{-2 w
\xi}}{1 - e^{-2w}} \right\},
\end{equation}
where $\xi = y/L$, $w = uL$. We have chosen the constant $c$ in
this equation so that
 $A(0) = A(L) = 0$, which  leads to Galilean coordinates
on both branes; one can easily see it from (\ref{A_sym}). From
this point of view the case under consideration can be called
"symmetric", although the function $A(y)$  (see Figure \ref{fig})
is not symmetric with respect to ${y}/{L}=0.5$.

The function $A(y)$ (\ref{fA}) is still rather complicated to
solve  the equation for the radion and gravity fields and find the
masses of the lowest and excited states. So we will again use the
approximation $uL\ll 1$, which proved to be quite reasonable.  In
this case we expand the exponential in $A(y)$ to the second order
in $2u|y| < 2uL \ll 1 $ and get
\begin{equation}\label{fAsym}
A(y) = \tilde k |y| +2b (uy)^2.
\end{equation}
Equation (\ref{sym}) in this approximation gives a relation among
the parameters of the model
\begin{equation}\label{relsym}
uL = -\frac{\tilde k}{2bu}.
\end{equation}
We also note that in this case the coordinates $\{x^\mu\}$ in
(\ref{metricDW}) are Galilean on both branes.

Assuming $b$ to be large enough, the integral relating the five-
and four-dimensional energy scales can be estimated as
\begin{equation}\label{intsym}
\int_{-L}^L e^{-2A(y)} dy \simeq \sqrt{\frac{\pi}{b}}
\frac{e^{(uL)^2 b}}{u}.
\end{equation}
The radion mass is again given by (\ref{mradap1}).  It turns out
that in this case it is possible to express the relation between
the five- and four-dimensional energy scales in terms of the
radion mass and the brane separation distance:
\begin{equation}\label{relsymrad}
M^2_{Pl}=4 \sqrt{2\pi} \frac{M^3}{m_{rad}}
e^{\frac{m_{rad}^2L^2}{32}}.
\end{equation}
If one  takes a rather small radion mass  $m_{rad} \sim 100
\,GeV$, so that the exponent in the (\ref{relsymrad}) is of the
order of unity, then one gets $M \sim 10^{12}-10^{13} GeV$.
 This  energy
scale arises, for example,  in the $SO(10)$  Grand Unification
model with the $SO(10)$ broken first to $SU(4)\times SU(2)\times
SU(2)$ as the intermediate  scale, at which the quark-lepton
$SU(4)$ symmetry is broken \cite{Langacker:1980js}.

There is also a possibility in this scenario to have the
fundamental five-dimensional energy scale of the order of $10
TeV$, but in this case the radion must be very heavy, although the
excitations of the tensor field remain in the TeV energy range.

Thus, we have shown that in the "bottom-up" approach, when the
bulk in the stabilized brane models is considered from the
viewpoint of an observer on "our" negative tension brane, the size
of the bulk   $L \sim TeV^{-1}$  appeared in a natural way. The
lowest radion state in the  most of the stabilized scenarios could
be rather light in the range of $100 GeV$, while its KK
excitations, as well as KK excitations of the graviton are
typically in the TeV range. In the most of the scenarios the
fundamental gravitational multidimensional scale is in the $TeV$
range. However, in the case of the "symmetric" warped scenario the
scale could be significantly larger in the range of $10^{12} -
10^{13} GeV$.

The existence of the stabilized warped bulk with $TeV^{-1}$ size
opens up new possibilities for studying models with KK excitations
of various bulk fields on "our" brane, which should naturally lie
in the mass range of $L^{-1} \sim 1 TeV$ leading to potentially
interesting collider phenomenology.

We would like to make one more remark. For certain values of the
parameters of the model the backreaction of the scalar field
essentially amounts to a renormalization of the parameter $k$ of
the RS1 model, which is related to the bulk cosmological constant
by $\Lambda = - 24M^3 k^2$ (\ref{cc}) and is responsible for the
generation  of the mass hierarchy. Explicitly, $k$ is replaced by
$\tilde k =(k-2bu)$ as in (\ref{fAap}). An interesting fact is
that the parameter $k$ is also renormalized in the unstabilized
RS1 model, when the Gauss-Bonnet term is added to the
gravitational action \cite{Kim:2000pz,Rizzo:2004rq}. For small
values of a dimensionless parameter $\alpha$, which stands in
front of the Gauss-Bonnet term together with the fundamental
energy scale $M$, the renormalization of $k$ looks like $k_{-} = k
- \alpha k^3/ M^2$. Depending on the sign of  $\alpha$, the
variation of $k$ can be either positive or negative, unlike only
the negative variation in the case of the stabilized model. It
would be interesting to find out, whether these two mechanisms
could compensate each other and produce a stabilized model with
the Gauss-Bonnet term and the Randall-Sundrum background metric
along the lines of paper \cite{Grzadkowski:2003fx}.

\bigskip

{ \large \bf Acknowledgments}

The work is partly supported by RFBR~04-02-16476,
RFBR~04-02-17448, Universities of Russia UR.02.02.503, and Russian
Ministry of Education and Science NS.1685.2003.2 grants. E.B. and
I.V.  would like to thank Bogdan Dobrescu, Tao Han, Chris Hill,
and  Jose Perez for clarifying discussions. E.B is grateful to the
Fermilab Theoretical  Physics Department for the kind hospitality.

\end{document}